# Shedding light on social learning

Kingsley J.A. Cox and Paul R. Adams
Department of Neurobiology, SUNY Stony Brook, NY, USA

**Abstract**

Culture involves the origination and transmission of ideas, but the conditions in which culture can emerge and evolve are unclear. We constructed and studied a highly simplified neural-network model of these processes. In this model ideas originate by individual learning from the environment and are transmitted by communication between individuals. Individuals (or "agents") comprise a single neuron which receives structured data from the environment via plastic synaptic connections. The data are generated in the simplest possible way: linear mixing of independently fluctuating sources and the goal of learning is to unmix the data. To make this problem tractable we assume that at least one of the sources fluctuates in a nonGaussian manner. Linear mixing creates structure in the data, and agents attempt to learn (from the data and possibly from other individuals) synaptic weights that will unmix - i.e. to "understand" the agent's world. For a variety of reasons even this goal can be difficult for a single agent to achieve; we studied one particular type of difficulty (created by imperfection in synaptic plasticity), though our conclusions should carry over to many other types of difficulty. We previously studied whether a small population of communicating agents, learning from each other, could more easily learn unmixing coefficients than isolated individuals, learning only from their environment. We found, unsurprisingly, that if agents learn indiscriminately from any other agent (whether or not they have learned good solutions), communication doesn't enhance understanding. Here we extend the model slightly, by allowing successful learners to be more effective teachers, and find that now a population of agents can learn more effectively than isolated individuals. We suggest that a key factor in the onset of culture might be the development of selective learning.

**Introduction**

Individual humans appear to solve problems by at least 2 synergistic mechanisms. First, they familiarize themselves with potentially relevant data and experiences, and, after thought, may experiment, tinker or make trial tests. These processes could be lumped together under the rubric of "discovery" or "invention". Second, they may copy or exploit, via supervised learning, communication and more formally education, the previous discoveries of others, adopting or adapting them to the task at hand. These mechanisms are synergistic in several ways. The details of many problems are unique, so even highly relevant prior art may need original modification. Prior discoveries were themselves built on even earlier ones, but may contain strikingly original elements. The gradual refinement and sharing of discoveries seems to underlie a process of "cultural evolution" (Boyd and Richerson, 1995). In particular, under conditions that are not fully understood, individual discoveries can be incorporated, via supervised learning from other individuals, into a shared body of knowledge, in a ratchet-like manner. Throughout this paper we assume that all forms of learning occur via synaptic plasticity mechanisms, while recognizing that genetic evolution is also a type of population-level learning process, and that synaptic and genetic learning may be coupled.

There are 2 main reasons why learning from others may be important. First, individual learning may be slow, compared to the lifetime of individuals. Second, individual learning may fail altogether, even with unlimited data. Slow or failed learning can reflect local minima or saddles in synaptic weight space (Atakulreka and Sutivong, 2007; Rattray, 2002; Wessels and Barnard, 1992; Saxe et al., 2013; Bengio, 2012), or unfavourable initialization (Bahri et al., 2020; Richards et al., 2019).

However, if everyone is both learner and teacher, and individual learning often fails or is incorrect, then learning from others may retard or completely prevent useful discovery (Cox and Adams, 2021; Rogers, 1988) rather than allowing cumulative, ratchet-like spread. For convenience we will refer to this rather obvious difficulty as the Rogers-Adams-Cox (RAC) dilemma, while recognising that the detailed mechanisms are quite different in the genetic and neural cases.

A core component of discovery is the recognition of significance or pattern in underlying complex, and initially seemingly random, data. These statistical regularities (e.g. higher order correlations; Simoncelli and Olhausen, 2001) often reveal the underlying processes generating the data. Indeed, one influential school of thought holds that to understand data, one must learn to generate it (Hinton, 2007).

There are 2 interrelated paradigms for detecting latent regularities in complex, apparently random, data. One is provided by the brain itself. Electrical signals in sets of neurons represent data, concepts etc, and these signals are successively transformed to more useful forms through banks of neurons linked by adjustable synaptic connections. The other is provided by statistics, artificial intelligence, and machine learning. The most successful example of this second paradigm is provided by artificial neural networks, which are loosely based on the first paradigm. Furthermore, recent advances in theoretical neuroscience draw upon progress in machine learning (Marques, 2021), which has provided insight into the difficulties that confront individual learners (Bengio, 2012).

We studied these issues using highly simplified models of both individual learning and communication. Although highly simplified these models still exhibit core features that hamper individual learning such as local minima and plateau. Input data are generated by linear mixing of independently fluctuating sources - the "independent component" (IC) paradigm (Bell and Sejnowski, 1997). Provided at least one of the sources has a nonGaussian distribution, and the data are suitably preprocessed, nonlinear Hebbian learning can easily find synaptic weights that allow unmixing (Hyvärinen et al., 2004). However, if starting weights or preprocessing is suboptimal and/or Hebbian adjustment isn't perfectly synapse-specific, individual learning can fail (Cox and Adams, 2009; Cox and Adams, 2014) To test whether communication with others can rescue individual learning, we constructed a model with multiple agents each of which can learn from others, using supervised learning, in addition to individual, unsupervised, learning. We previously reported that because agents can learn from other agents which have found incorrect solutions, communication per se doesn't allow the spread of discovery (RAC dilemma) (Cox and Adams, 2021). We suggested that, for ratchet-like cultural evolution, successful learners should be privileged teachers. Here we test this hypothesis, by slightly modifying that model to incorporate an additional factor, which we call "light", which represents the degree to which supervised learning is accelerated when the supervising agent reaches or nears a correct solution.
It's important to note that while our "toy" model assumes extremely simplified generative statistics, the underlying problem, the RAC dilemma, is likely to be even more severe for more realistic assumptions, and our proposed general solution, "light", is likely to be equally relevant.

**Methods**

We studied the behavior of a small (n = 2 or 4) set of interacting 1-neuron "agents". Each agent had 2 variable-weight inputs, and could learn (activity-dependent weight adjustment) either unsupervised by Hebbian weight adjustment triggered by current input values or supervised using a standard Delta Rule, comparing the agent's current output in response to the current input vector to the output of a "teacher" agent which is also attempting to learn the correct solution. The goal of learning is to find synaptic weights that permit agents to track the current value of a hidden randomly-fluctuating (but nonGaussian) cause or "source" (the "IC") that has been linearly mixed with another confounding Gaussianly-fluctuating source. In the simplest case individual agents can always correctly learn unmixing weights (Hyvärinen et al, 2004; Hyvärinen and Oja, 1998) but our model adds complicating factors which can prevent successful individually learning - either imperfect preprocessing of input data (imperfect "whitening") or imperfect Hebbian learning (minor synapse-inspecificity Cox and Adams, 2014; Elliott, 2012). This paradigm is mathematically transparent but nevertheless presents the features that make individual learning difficult or impossible (learning slowed by saddle-point trapping, false minima, suboptimal starting weights etc.). We are interested in the crucial issue of whether social learning - instruction by already successful agents - can rescue difficult individual learning even when some agents are unsuccessful, allowing the onset of "cultural evolution".

Our previous minimal model of coupled social and individual learning (Cox and Adams, 2021) was based on earlier results (Cox and Adams, 2009; Cox and Adams, 2014) with single-agent learning in the ICA model when learning was hampered by connection update inspecificity ("crosstalk") and/or inadequate preprocessing. In that model the same input data **x** (generated by linear mixing of independently fluctuating "sources", **x** = **Ms**) were seen by all agents. In the present study we modified the supervised learning rule by incorporating a factor we call "light".

**Unsupervised learning with ICA**

The goal in ICA is to unmix a set of independent unit-variance sources that have been mixed by a mixing matrix

**x** = **Ms**

where **x** is the observable vector of mixed inputs, **s** is a vector of hidden independent sources, and **M** is the mixing matrix. In all our simulations **M** was orthogonal and there were only two randomly fluctuating sources, one with a Gaussian distribution and the other Laplacian. This ensures that, in the absence of crosstalk, there is only one learnable solution, a synaptic weight vector pointing in the same direction as the column of the mixing matrix associated with the nonGaussian source (Rattray, 2002). For simplicity we only consider one output neuron which tries to find a weight vector that will 'unmix' the sources so that the output neuron tracks one of them.

We use the one unit negentropy-maximizing rule (Hyvärinen and Oja, 1998)

$\Delta$**w** = k σ **x** f(**w**$^T$ **x**), normalize **w**

where k is a learning rate, **w** is the weight vector, y is the output, and σ is either +1 or -1 depending on the nonlinearity and the nonGaussian source statistics (Cichocki et al., 1997). In our case, with n=2, weight vector normalization after each iteration means that the 'basin' of attraction of the learning dynamics lies on the unit circle (which due to symmetry can be contracted to an arc).

Thus the output neuron sums the inputs weighted by the current weight vector **w** and then passes this total 'activation' through a nonlinearity f(y). In this paper we explored both cubic ($y^3$) or tanh nonlinearities (Hyvärinen et al., 2004). The current weight vector is then increased by an amount proportional to f(y) and finally the updated weight vector is normalized so that no individual weight can become too large. **M** is constrained to be orthogonal so that there are no pairwise correlations introduced into the input vectors **x** (i.e. the source vectors are only rotated). Even if the original mixing is nonorthogonal this condition can be achieved by suitable preprocessing. The weight vector that enables y to track the nonGaussian source is called the independent component (IC).

We only used 2 sources, however the single Gaussian source can be replaced by multiple Gaussian sources with the same outcome (Elliott, 2012).

We denote unsupervised learning as 'U'.

**Crosstalk**

Inspecificity in the update rule is introduced by modifying the update in **w** by using an error matrix **E** as described in (Cox and Adams, 2009; Cox and Adams, 2014; Radulescu et al., 2009)

$$\mathbf{E} = \begin{bmatrix} 1-e & e \\ e & 1-e \end{bmatrix}$$

where *e* is the, typically very low, crosstalk level. Neural network models usually assume perfectly connection-specific weight adjustments, but this isn't possible in real brains, because synapses are extremely close-packed (Rădulescu et al., 2009).

**The mixing matrix**

The mixing matrix is orthogonal and can be varied by changing the angle θ that the first column of **M** makes with the vector $[1,0]^T$. In this paper $\theta$=45

$$\mathbf{M} = \begin{bmatrix} cos\theta & -sin\theta \\ sin\theta & cos\theta \end{bmatrix}$$

**Principal component of E**

Because M is orthogonal the data is white, and with crosstalk, an eigenvector of **E** becomes a possible fixed point of the dynamics in addition to the IC (Elliott 2012; CA 2014). With non-zero *e*, the eigenvectors of **E** are the vectors [0.707, 0.707] and [0.707, -0.707] regardless of the amount of error (since **E** is a symmetric matrix). When using a tanh nonlinearity σ is -1, and with crosstalk the PC associated with the least eigenvalue is the stable fixed point of the dynamics [0.707, -0.707] and we call this the 'PC'. With the cubic nonlinearity σ is +1 and the PC associated with the largest eigenvalue is the stable fixed point i.e. [0.707, 0.707]. (Refer to (Cox and Adams, 2021) for details).

**Basins and fixed points**

The results obtained in previous work (Cox and Adams, 2009; Cox and Adams, 2014) were discussed in terms of the stable fixed points (IC and PC) and the same is done in this paper. Figure 1 below illustrates basins for the IC and PC with the fixed points being at the end of the red and blue lines. Agents move along the line to the fixed points. With supervised learning (see below), however, an agent, say the 'blue' one in the PC basin, can move in the opposite direction, towards the 'teacher's' basin. The black dot between the red and blue lines represents the separatrix between the basins.

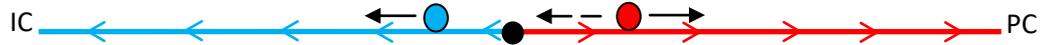

Figure 1  Illustration of dynamics of an agent.  Each red or blue line segment represents the weights that would result in the agent moving towards the corresponding fixed point with the red and blue arrows showing the direction of movement. When learning unsupervised each agent moves as shown with the solid black arrows, but when one agent, the blue one in the PC basin for instance, learns from the other agent (in the IC basin) it moves towards the IC basin as shown with the dotted arrow.

**Supervised learning**
We define an agent as a single output neuron feed-forward neural network that is trying to solve the ICA problem, i.e. tracking the nonGaussian source of the input data.  It can do this either directly from observing input fluctuations and using unsupervised Hebbian learning or by supervised learning (denoted by 'S'), i.e. by also observing the current output of another agent that may have already learned the correct solution, in addition to directly observing the input vectors.
For supervised learning, agents use the Delta rule (Widrow and Hoff, 1960). The difference between the agent's output y is compared against a teacher's output $y_T$ (in this case another agent's estimate of the current value of the nonGaussian source). In this case the output y is the dot product of the input vector and the weight vector and has not been put through a nonlinearity.

$$\Delta \mathbf{w} = k\, \mathbf{x}(y - y_T)$$

where $\Delta\mathbf{w}$ is the update in the weight vector and k is a learning rate. Crosstalk slows learning for the supervised agent but does not change the direction of its weight vector (Cox and Adams, 2021).

**Interacting agents**
We studied situations in which 2 or 4 potentially interacting agents can randomly switch their learning strategies between unsupervised and supervised, while all observing the same input vectors. At each iteration each agent randomly decides if it will learn U or S.
With 2 agents each agent learns on its own (U) or from the other agent (S).
With 4 agents each agent either learns U, or S from an average of the other agents' outputs.
Each agent chose randomly between supervised and unsupervised learning: the supervised learning used the Delta Rule as before but this time an average of the outputs of the other agents was used instead of the output of a randomly chosen individual agent. This "meanfield" simplification gave similar results to runs under otherwise identical conditions where an agent learned from only one randomly chosen other agent.

**Light**

To test the idea that selective learning from privileged teachers could overcome the RAC dilemma, we added to the foregoing framework a feature we call 'light', so that in the delta rule the learning rate is multiplied by a function of the cosine of the current angle, ϕ, of the teacher's weight vector with the IC, a factor we call 'light'. "Light" could reflect a variety of psychological or social processes, some of which we will briefly discuss. However the key aspect captured by our extremely simplified mathematical formulation is that selective learning from teachers who have themselves successfully learned could allow the onset of a cultural ratchet, whereby a population of learners becomes gradually more successful, even though a population of non-communicating agents fails to learn (because they mostly start in the PC basin). There are 2 parameters that can be set to change the amount of light, a magnitude factor *a* and a steepness factor *b* as follows

light = $a(\cos(\phi))^b$

Note that in this formulation teachers become more effective the closer they are to the correct solution. Of course if only the agents that have learned exactly correct solutions are allowed to teach, the RAC dilemma evaporates, at least in the large population limit.

# Results

**2 agents no light**

First, we look at the behaviour of 2 agents, one starting at the PC and one at the IC, when the crosstalk is at a level (0.07) which splits the PC/IC basin in halves (Cox and Adams, 2014; Cox and Adams, 2021) which will serve as a basis for later results with 'light'. As we have shown previously (Cox and Adams, 2021) both agents either end up at the PC or the IC depending on the actual particular noisy (because the source signals are themselves random) movements of the agents' weight vectors (learning U or S). Because each agent starts the same distance away from the separatrix both end up very close to the PC 50% of the time and 50% both end up very close to the IC. Figure 2 shows examples of individual runs. In the first one (panel a) both agents happen to arrive near the separatrix (green horizontal line) at the same time and both linger there for some time before the 'IC' (blue) agent (i.e. the one that starts closest to the IC) slips over into the PC basin and then both agents move together down the basin to the PC. In the second run (panel b), under identical conditions, the IC agent happens to move a bit faster towards the separatrix, some time before the PC agent has reached it, and then both agents move faster to the PC. In the third and fourth runs (panels c and d) the PC agent reaches the separatrix first and both agents end up at the IC.

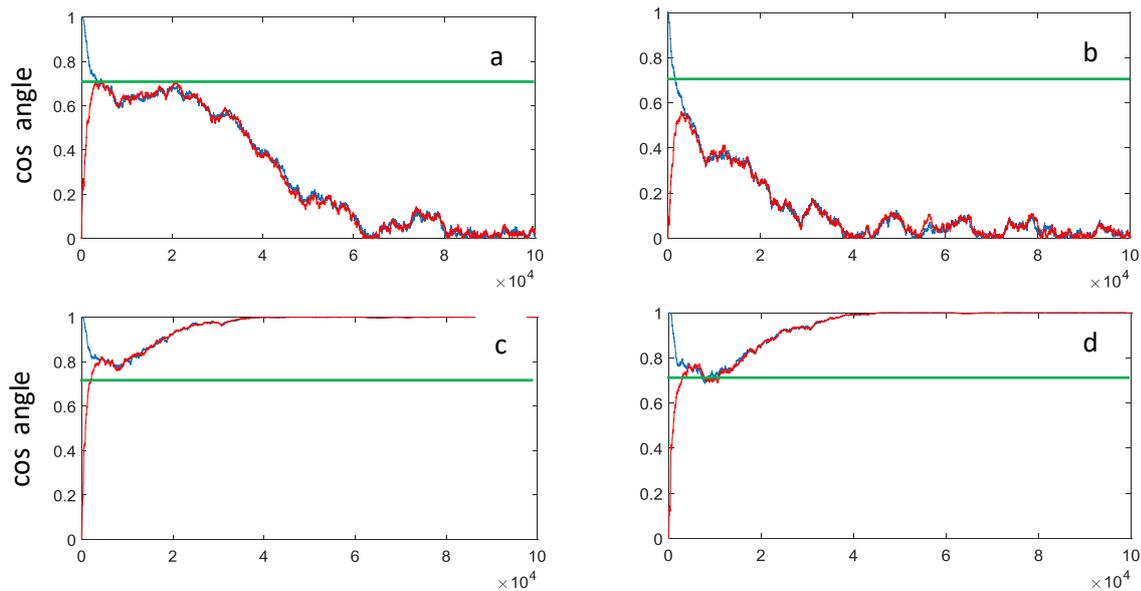

Figure 2 Dynamics of learning with 2 agents. The y axis is the cosine of the angle between the agent's weight vector and the IC direction, and iterations are along the x axis.

**With Light (tanh nonlinearity)**

In the absence of light, at a crosstalk level of 0.08, (as opposed to 0.07) the PC basin is larger than the IC basin and both agents always go to the PC (figure 3 left panel). However, with light ($a$ and $b$=1, see Methods) both end up at the IC almost all of the time (figure 3 right panel). Figure 3 shows the trajectory of each agent in both cases.

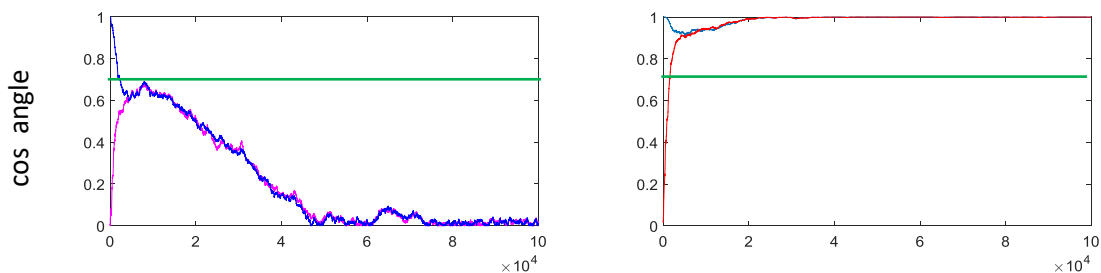

Figure 3 Dynamics of learning with 2 agents and 'light' using a tanh nonlinearity. 2 agents, one starting from [-1,1] (PC, red) and one from [1,1] (IC, blue), crosstalk at 0.08 and LR 0.001. Cosine of angle of weight vectors of agents with the IC are plotted. The left hand plot is without light and the right hand plot with light. Without light at this crosstalk level both agents always go to the PC, but with light ($a$ and $b$=1) they both go to the IC.

**2 agents tanh – average behaviour over many runs**

For 2 agents using the tanh rule, starting one at the PC and one at the IC, with crosstalk = 0.08 (greater than the threshold value of 0.07 to ensure a bias to the PC), without light 100% of the time they both end up at the PC. With light ($a$ and $b$ = 1) this changes to only 1.6% of the time both ending up at the PC – they almost always go to the IC now (figure 4)

This was then repeated with the cubic rule with slightly different results. For light with *a* = *b* = 1 the shift is less pronounced, moving the crossover point from 0.14 to 0.15 (for the cubic rule with 0.14 crosstalk half the time both agents go to the PC and half the time to the IC). Increasing the amount of light shifts the crossover further. With *a*=2 and *b*=2 the crossover point (see below) moves from 0.14 to 0.19 (figure 4).

Note that in different runs using the same crosstalk and mixing matrix both agents always move to the same fixed point because of the synchronising effect of supervised learning, and at low crosstalk the fixed point is close to the IC and at high crosstalk the fixed point is close to the PC. However, at intermediate crosstalk values, and finite learning, in different runs either the IC or PC can be chosen, depending on the particular sequence of random input vectors (figs 4-6). The graphs show the proportions going either to the IC or to the PC in repeated different runs in this intermediate range. We refer to the (interpolated) crosstalk value at which the IC or PC are chosen with equal probability as the "crossover" value.

Previously we had shown that whether or not both agents both go to the PC or to the IC depends on the crosstalk level, and that the crossover value going to the PC or IC was 0.07. When light was added this threshold changed to 0.1 which indicates that a higher error was tolerated for a 50:50 basin split. Thus addition of light favours correct learning.

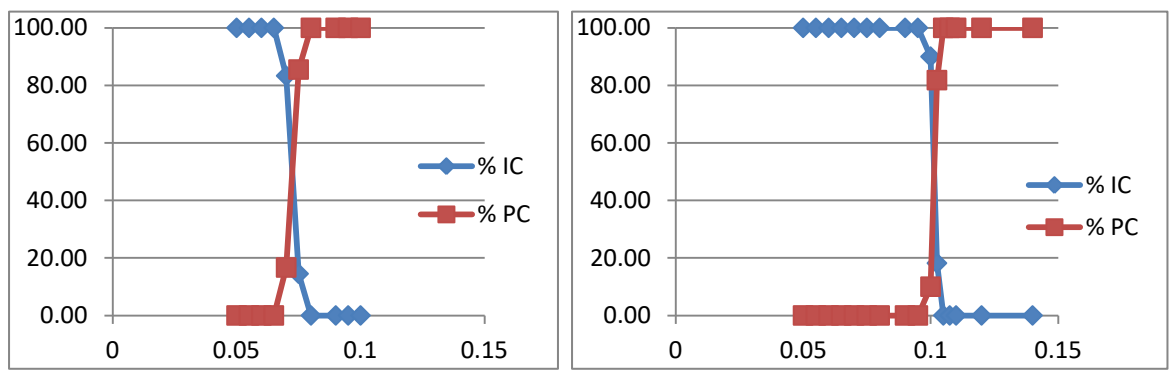

Figure 4  Fate of interacting agents as crosstalk is varied with tanh nonlinearity. For multiple runs one agent was started at the IC and the other agent at the PC. The plots show the percent of time both agents went to the IC or PC at different crosstalk levels. Each run was for 200,000 iterations and at each iteration both agents could change learning type (i.e. if say agent 1 was currently U then it would randomly pick whether to do U or S for the next iteration). The left hand graph shows how learning progressed without light, and the right hand graph has light. The y-axis shows the % of the time both go to either the IC or PC.  Note how the crossover point shifts with light from 0.07 to 0.1

A similar result was obtained using a cubic nonlinearity, except that stronger light was required, as shown below in figure 5.

## Cubic nonlinearity

Similar result were obtained with a cubic nonlinearity as shown in figure 5 for 2 agents

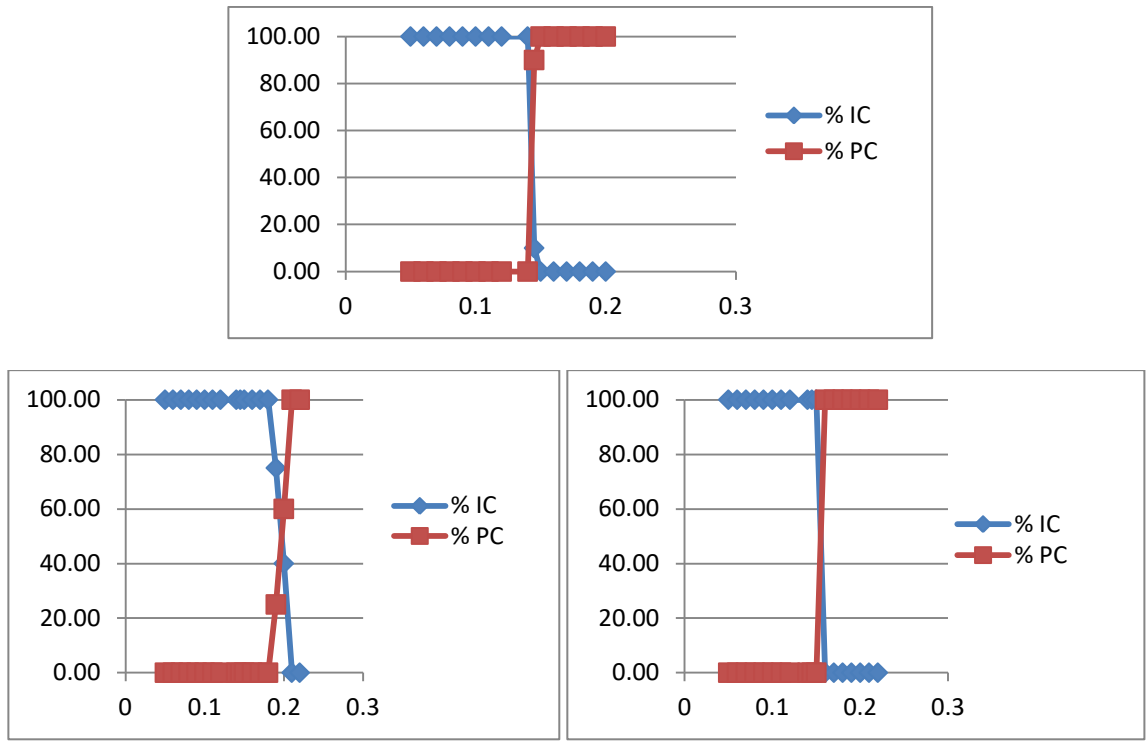

Figure 5 Cubic 2 agents Top panel no light with crossover point at 0.14. Bottom left panel basic light showing crossover point at 0.15 and bottom right hand panel LT=2 and power = 2 with crossover point at 0.19

## 4 agents

With 4 agents, 2 starting at the PC and 2 starting at the IC, with no light crossover occurs at 0.14 (as in the 2 agent case). With light (in this case 'stronger' light with a=2 and b=2) crossover shifts to 0.16. See figure 6.

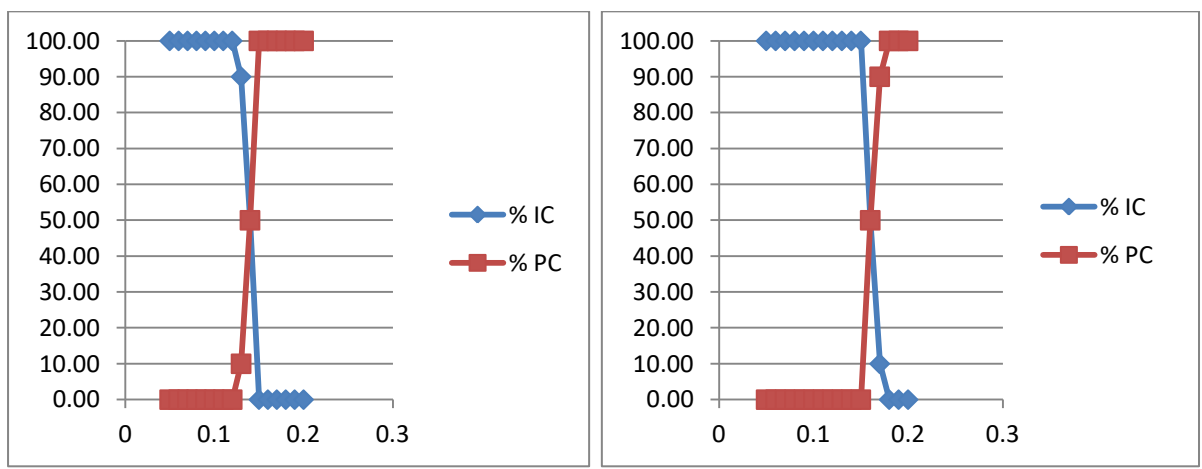

Figure 6 Cubic nonlinearity, 4 agents. Left hand panel no light and right hand panel LT=2 and power = 2. The crossover point moves from 0.14 to 0.16

It is interesting to see the trajectory of learning with 4 agents with and without light, with two starting close to the PC and two starting close to the IC (figure 7).

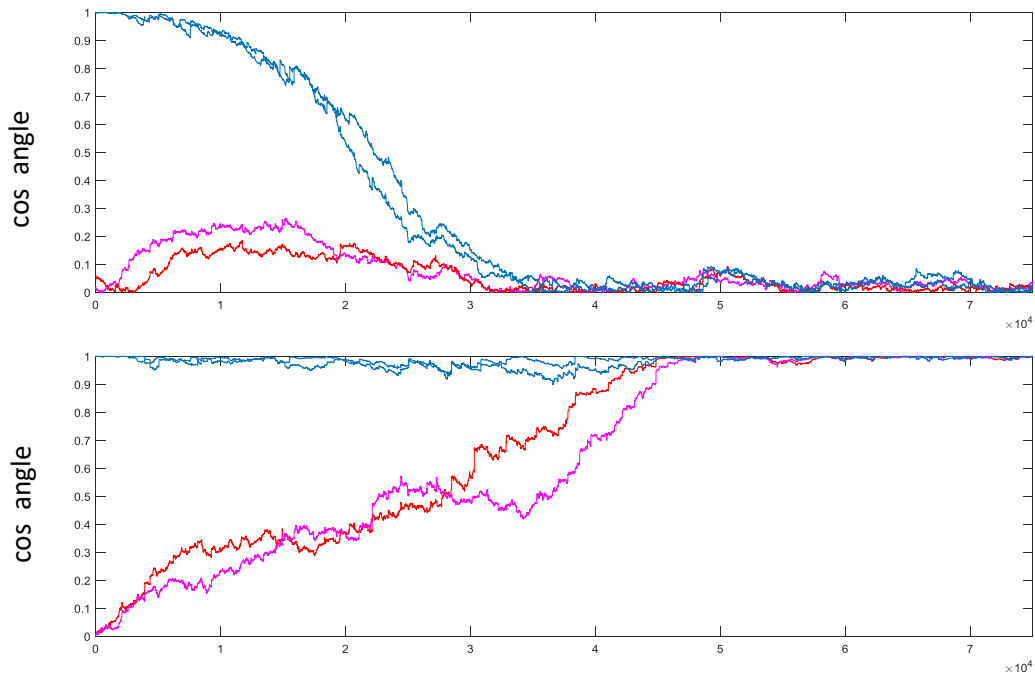

Figure 7  Dynamics of learning with 4 interacting agents. Here 2 agents (blue lines) start at [1,1] i.e. the PC, and 2 at [-1,1] i.e. the IC (red lines), crosstalk of 0.15 and cubic nonlinearity.  The cosine of the angle between each agent's weight vector and the IC is plotted (y axis with iterations along x axis). In the run without light (top panel) the 2 PC agents (red) drag the 2 IC agents (blue) over to the PC basin. In another run with light (bottom panel) the 2 IC agents (blue) drag the 2 PC agents (red) over to the IC basin. In the top panel the agents at the PC move a little from the PC as the agents near the IC start to move – there is a 'struggle' as the agents learn from each other with in this case the PC agents winning. The reverse is true in the bottom panel where the IC agents are aided by 'light'.

**4 agents 3 PC 1 IC**

The simulations above were all done with an equal number of agents starting in each basin. We also explored the situation where one pioneer agent began from the IC basin and three from the PC basin in order to see if it was possible for a single IC agent with light to attract the three PC agents to the IC basin. The threshold was found for the amount of error for all agents to go to the PC all the time (0.05), and then light was added. With light (a=2) it was found that the interpolated crossover point was moved to 0.08. See figure 8.

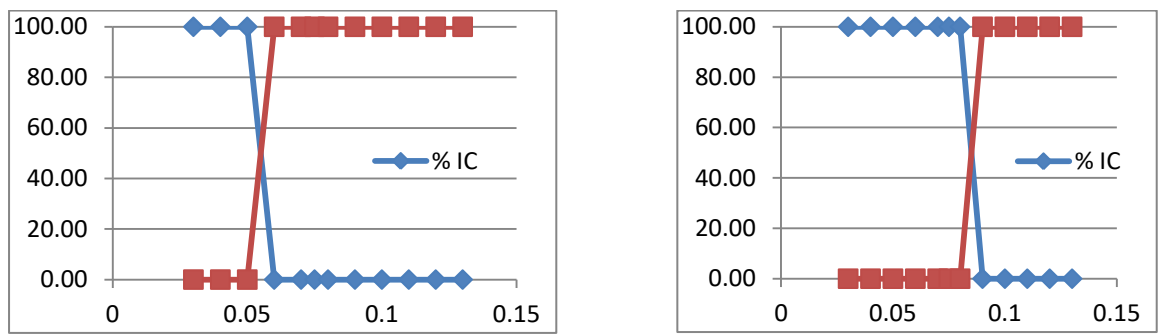

Figure 8  3 agents at PC and 1 at IC.  The left hand panel shows the behaviour of the system when 3 agents begin at the PC and 1 at the IC (tanh) and it can be seen that at an error rate of 0.04 and below all agents go to the IC basin. The right hand panel shows the same system but this time with light (a=2) and under these conditions the crossover point shifts to 0.08.

With three agents starting in the PC basin the error level at which all agents converge to the PC basin is 0.05 (this is different to the case where equal numbers of agents start in the PC and IC basins since there each agent is learning from more agents already in the PC basin). However, with light the single agent in the IC basin is able to drag over the 3 agents in the PC basin even up to an error level of 0.08. Above 0.08 all agents end up at the PC at this light level (a=2).

In summary, the introduction of light always allows the population of interacting agents to more reliably find correct solutions under more difficult conditions (greater crosstalk) than in the absence of light. Of course without interaction agents starting at incorrect solutions (or in basins leading to incorrect solutions) will always stay there.

**Discussion**

Understanding our world, and identifying the hidden causes of the endless stream of data that reaches our senses, is useful for survival and reproduction. The basic machinery of this learning process seems to be Hebbian synaptic plasticity since this confers sensitivity to input correlations. But because both the data and the underlying causal mechanisms are often subtle and complex, understanding can be slow and incomplete - and often dies with the individual animal. Many authors have proposed that humans can achieve deeper understanding by standing on the shoulders of other humans (Newton, 1675), in a cultural "ratchet" whereby useful ideas are invented, shared, tested and refined by members of communicating populations. We seek to better understand this process by using extremely simplified formal models of the way the sensory data are generated and the brain learns about their statistical structure. We use a neural network framework, and the simplest possible causative framework, a linear ICA model (Bell and Sejnowski, 1997; Hyvärinen et al., 2004). Indeed we reduce a whole brain to a single "neuron" equipped with just 2 sensory inputs arriving via 2 synapses endowed with Hebbian synaptic plasticity. We assume that brains/agents can interact, via an unspecified but accurate and presumably symbolic process, to inform each other about their tentative inferences. If the individual learning process is straightforward, with only 1, correct, fixed point of the learning dynamics, all agents can individually unsupervisedly learn correct solutions, and since they learn faster if they happen to start close to the correct solution, communication together with supervised learning, rather trivially, accelerates learning at the population level (Cox and Adams, 2021).

We then move slightly closer to reality by assuming that individual learning is sometimes unsuccessful. Specifically in the neural/ICA framework we assume that the plasticity process may not be completely synapse-specific and/or that the data have not been adequately preprocessed, so that agents can get trapped in false minima (i.e. incorrect solutions). Nevertheless this model remains, in the individual agent case, mathematically transparent (Elliott, 2012).

In our previous study in this framework (Cox and Adams, 2021) we showed that when all agents can learn both socially and individually, and individual learning is hampered by plasticity crosstalk, social learning doesn't help. In particular, individual learners can arrive at either correct or incorrect solutions, depending on their starting weights and intrinsic noise in the learning process. We found that, as expected, social learning does synchronize learning, in the sense that all learners tend to learn the same thing, especially when the noise is low. However, even when one or more of the agents starts at the correct solution all of them move to the wrong solution when crosstalk reaches a critical level (see figure 8 in (Cox and Adams, 2021) and figures 4,5 and 6 in this paper.) Although in the paper we explored only one particular form of learning difficulty, synaptic crosstalk, it is very likely that the beneficial effects of light would apply also to the closely related case of imperfect data whitening (Cox and Adams, 2014). This could extend more generally to all types of learning difficulty (false minima, plateau etc.)

This result is reminiscent of "Rogers' Paradox" (Rogers, 1988), which shows in an evolutionary game model that the overall fitness of a population isn't affected by social learning. In both cases the underlying reason is that social learning is only useful if individual learning about the environment is correct. We suggested that this difficulty, which we dub the RAC dilemma, could be overcome if those individuals who have successfully learned useful solutions become more effective teachers. In the present study we tested this idea by incorporating a factor we call "light" into the supervised learning rule. This "light" factor represents how close an agent currently is to the correct solution. Our model is agnostic as to the various actual processes which could underlie "light". Light could reflect prestige, success, trust, expertise, science or an otherwise privileged status, or any combination of such attributes. Our only assumption is that teaching is more effective when teachers have learned something valid about the current environment. Our key result is that the addition of "light" increases the crosstalk level that forces all agents to the wrong solution, even though at least one agent is initially already at the correct solution. In other words, the addition of light allows the whole population to learn the correct solution in conditions in which otherwise no correct learning would be possible.

In essence light is a meta-form of supervised learning, which one could call "super-supervised" learning. In regular supervised learning it's assumed there is a perfect "teacher", an oracle, providing the target output(s) which guides learning. But in the present model all agents act as teachers as well as learners, and if agents are not already at the correct solution, correct learning can completely stall. If agents learn faster from agents that are closer to (or already at) the correct solution, by the addition of light, this failure can be prevented. While social learning uses the delta rule, without light it's not "supervised" in the traditional sense since no oracle is available. Oracles only become available as a result of successful individual, unsupervised, learning, which is typically slow, unreliable and often ineffective. In our model, the "super-supervision" arises as the result of unidentified social processes which generates "light". (However presumably if all agents started at the correct solution, they would presumably stay there.)

It might be argued that our assumptions and formalism are unrealistic because they fail to embody the fact that if an agent "knows" whether it has achieved success it could in principle use that knowledge to accelerate its own learning. In our model an agent that is learning has access to the cosine of the angle (relative to the IC) a teacher sits at, but the teacher itself doesn't have access to the cosine (which would allow it to use traditional supervised learning). Social "supervised" learning is non-traditional because the teachers are as blind as the learners (and indeed are the same agents). We believe this condition inevitably controls the efficacy of social learning. Learning is only supervised in the sense that a delta rule is employed, not in the sense that an oracle guides learning.

"Weakly supervised" learning refers to a situation where labelled data drive learning, but the labels are noisy (Zhou, 2017). Our model is clearly related to this approach, but nevertheless distinct, since the "noise" in the labels is not random but reflects the difficulty of the individual learning problem (especially the dependence of successful learning on optimal initial weights).

In summary, we investigated a highly simplified model of a way in which social learning could overcome difficulties that isolated individual learners face. We cannot at this stage speculate about the many possible and likely synergistic factors that might actually underlie the general factor we here call "light".